\documentclass[letterpaper]{article}
\usepackage{proceed2e}
\usepackage[margin=1in]{geometry}

\usepackage{times}

\usepackage{amsbsy}
\usepackage{amsmath}
\usepackage{amsfonts}
\usepackage{amsthm}
\usepackage{amssymb}
\usepackage{algorithmic}
\usepackage{algorithm}
\usepackage{enumerate}
\usepackage{graphicx}
\usepackage{multirow}
\usepackage{natbib}
\setcitestyle{round}
\usepackage{subfigure}
\usepackage{listings}
\usepackage{xcolor}
\theoremstyle{plain}
\newtheorem{theorem}{Theorem}

\theoremstyle{definition}
\newtheorem{experiment}{Model} 

\theoremstyle{remark}

\title{Independent Component Analysis via Energy-based and Kernel-based Mutual Dependence Measures}

\author{ {\bf Ze Jin\thanks{\, Corresponding author. Email address: zj58@cornell.edu.}} \\
Department of Statistical Science \\
Cornell University\\
Ithaca, NY 14850 \\
\And
{\bf David S. Matteson\thanks{\, Research support from an NSF Award (DMS-1455172), a Xerox PARC Faculty Research Award,
  and Cornell University Atkinson Center for a Sustainable Future (AVF-2017).}}   \\
Department of Statistical Science\\
Cornell University \\
Ithaca, NY 14850    \\
\\
}

\begin{document}

\maketitle

\begin{abstract}
We apply both distance-based \citep{jin2017generalizing} and kernel-based \citep{pfister2016kernel}
mutual dependence measures to independent component analysis (ICA),
and generalize dCovICA \citep{matteson2017independent} to MDMICA,
minimizing empirical dependence measures as an objective function in both deflation and parallel manners.
Solving this minimization problem, we introduce Latin hypercube sampling (LHS) \citep{mckay2000comparison},
and a global optimization method, Bayesian optimization (BO) \citep{mockus1994application}
to improve the initialization of the Newton-type local optimization method.
The performance of MDMICA is evaluated in various simulation studies and an image data example.
When the ICA model is correct,
MDMICA achieves competitive results compared to existing approaches.
When the ICA model is misspecified,
the estimated independent components are less mutually dependent than the observed components using MDMICA,
while they are prone to be even more mutually dependent than the observed components using other approaches.
\end{abstract}


\section{INTRODUCTION}\label{intro}

Since most natural processes have multiple components,
multivariate analysis is more compelling than univariate analysis.
Nevertheless, multivariate analysis is considerably more complicated than univariate analysis,
because it accounts for the mutual dependence between all variables.
Due to the curse of dimensionality, it becomes essential to interpret multivariate data
through a simplified representation via dimension reduction.

Independent component analysis (ICA) represents multivariate data by mutually independent components (ICs).
Thus, linear combinations of ICs capture the structure of multivariate data
even when other linear projection methods, such as principal component analysis (PCA), are not sufficient.
As a classical unsupervised learning method, ICA has been developed for applications including
blind source separation, feature extraction, brain imaging, etc.
\citet{hyvarinen2004independent} provide a comprehensive overview of ICA approaches to estimate ICs.

Let $Y = (Y_1, \dots, Y_d)' \in \mathbb{R}^d$ be a random vector as observations.
Assume that $Y$ has a nonsingular, continuous distribution $F_Y$, with $\textrm{E}(Y_j) = 0$
and $\textrm{Var}(Y_j) < \infty$, $j = 1, \dots, d$.
Let $X = (X_1, \dots, X_d)' \in \mathbb{R}^d$ be a random vector as ICs.
In particular, the univariate components $X_1, \dots, X_d$ are mutually independent,
and at most one component $X_j$ is Gaussian.
Without loss of generality, $X$ is assumed to be standardized
such that $\textrm{E}(X_j) = 0$ and $\textrm{Var}(X_j) = 1$, $j = 1, \dots, d$.
A linear latent factor model to estimate $X$ from $Y$ is given by
\begin{equation*}\label{ica1}
Y = MX,
\end{equation*}
where $M \in \mathbb{R}^{d \times d}$ is a nonsingular mixing matrix.

Prewhitened random variables are uncorrelated and thus more convenient to work with
from both practical and theoretical perspectives.
Let $\Sigma_Y = \textrm{Cov}(Y)$ be the covariance matrix of $Y$,
and $H = \Sigma_Y^{-1/2}$ be an uncorrelating matrix.
Let $Z = HY = (Z_1, \dots, Z_d)' \in \mathbb{R}^d$ be a random vector as uncorrelated observations,
such that $\Sigma_Z = \textrm{Cov}(Z) = I_d$, the $d \times d$ identity matrix.
Then the relation between $Z$ and $X$ is
\begin{equation}\label{ica2}
X = M^{-1}Y = M^{-1}H^{-1}Z \triangleq WZ,
\end{equation}
where $W = M^{-1}H^{-1} \in \mathbb{R}^{d \times d}$ is a nonsingular unmixing matrix.
Given that $Z$ are uncorrelated, $W$ is an orthogonal matrix, with $d(d - 1)/2$ free elements rather than $d^2$.
We aim to simultaneously estimate $W$ and $X$,
such that the components of $X$ satisfy the assumption of mutual independence.

Many popular ICA approaches
minimize the mutual information or maximize the non-Gaussianity
of the estimated components under the constraint that they are uncorrelated.
Examples include the fourth-moment matrix diagonalization of FOBI \citep{cardoso1989source}
and JADE \citep{cardoso1993blind},
the information criterion of Infomax \citep{bell1995information},
the maximum negentropy of FastICA \citep{hyvarinen1997fast},
and the maximum likelihood principle of ProDenICA \citep{hastie2003independent} and Spline-LCA \citep{risk2016linear, jin2017optimization}.

Some other ICA approaches minimize the mutual dependence between the estimated components using a specific dependence measure.
While dependence measures have been extensively studied,
two classes have attracted a great deal of attention.
One is the distance-based energy statistics \citep{szekely2013energy}.
\citet{szekely2007measuring} proposed distance covariance (dCov) to measure pairwise dependence,
and \citet{jin2017generalizing} extended it to mutual dependence measures (MDMs).
Another is the kernel-based maximum mean discrepancies (MMDs) \citep{gretton2007kernel}.
\citet{gretton2005kernel} proposed Hilbert$-$Schmidt
independence criterion (HSIC) to measure pairwise dependence,
and \citet{pfister2016kernel} generalized it to $d$-variable Hilbert$-$Schmidt
independence criterion (dHSIC) measuring mutual dependence.
\citet{sejdinovic2013equivalence} showed that these two classes of measures
are equivalent in the sense that
MMDs can be interpreted as energy statistics with a distance kernel,
and energy statistics can be interpreted as MMDs with a negative-type semimetric.

Meanwhile, \citet{chen2005consistent} and \citet{eriksson2003characteristic}
applied a characteristic function-based dependence measure to ICA,
for which \citet{jin2017generalizing}
provided a closed-form expression as an MDM and studied its asymptotic properties.
\citet{bach2002kernel} applied a kernel-based dependence measure to ICA,
which was formulated as an HSIC in \citet{gretton2005kernel}.
Motivated by the properties of HSIC,
\citet{shen2007fast} proposed FastKICA based on a mutual dependence measure extension,
which is the sum of all pairwise HSIC while its 0 value does not imply mutual independence.
Inspired by the properties of dCov,
\citet{matteson2017independent} proposed dCovICA based on another mutual dependence measure extension,
which is a sum of squared dCov and equals 0 if and only if mutual independence holds.

However, \citet{matteson2017independent} only demonstrated the results of a single measure from the class of energy-statistics,
using multiple values to initialize the local optimization without any comparison.
Thus, in this paper, we generalize dCovICA to a new approach, MDMICA,
by applying the mutual dependence measures proposed in \citet{jin2017generalizing} and \citet{pfister2016kernel},
and make two contributions as follows.
First, we extend its ICA framework to accommodate mutual dependence measures from both classes of energy statistics and MMDs,
and compare the performance of these measures in numerical studies.
Second, we study the non-convex optimization problem when estimating ICs under this ICA framework,
and investigate the improvement of using multiple values over a single value for initialization
through Latin hypercube sampling, a random sampling method.
In addition, we introduce a global optimization method, Bayesian optimization,
to further improve the initialization of local optimization.

The rest of this paper is organized as follows.
We generalize the ICA framework of dCovICA in Section \ref{ica}.
In Section \ref{mdmica}, we give a brief overview of dCov and MDMs,
propose the new ICA approach, MDMICA, based on MDMs,
and derive its asymptotic properties.
In Section \ref{global}, we introduce Latin hypercube sampling
and Bayesian optimization to aid the initialization of subsequent
local optimization method when estimating ICs.
We present the simulation results in Section \ref{sim}, and a real data example in Section
\ref{data}\footnote{See CRAN for an accompanying R package \texttt{EDMeasure}.}. 
Finally, Section \ref{con} summarizes our work.

\section{ICA FRAMEWORK}\label{ica}

For $d \geq 2$, the group of $d \times d$ orthogonal matrices is denoted by $\mathcal{O}(d)$, and
its subgroup with determinant 1 is denoted by $\mathcal{SO}(d)$.
For $i \neq j$, we start with the identity matrix $I_d$, and substitute $\cos(\psi)$ for the $(i, i)$ and $(j, j)$ elements,
$-\sin(\psi)$ for the $(i, j)$ element, and $\sin(\psi)$ for the $(j, i)$ element,
then we obtain a Givens rotation matrix denoted by $Q_{i, j}(\psi)$.

Let $\theta = \{\theta_{i, j}: 1 \leq i < j \leq d \}$
denote a vector of rotation angles with length $p = d(d - 1)/2$,
and let $\theta_i = \{\theta_{i, j}: i + 1 \leq j \leq d \}$ such that $\theta = \{\theta_{i}: 1 \leq i \leq d - 1\}$.
Then any rotation matrix $W \in \mathcal{SO}(d)$ can be parameterized via $\theta$ as $W(\theta)$,
or equivalently a product of $p$ Givens rotation matrices determined by $\theta$ as
\begin{equation*}
W(\theta) = G^{(d-1)}(\theta_{d-1}) \dots G^{(1)}(\theta_1),
\end{equation*}
where $G^{(k)}(\theta_k) = G_{k,d}(\theta_{k,d}) \dots G_{k,k+1}(\theta_{k,k+1}) $
represents the rotations of the $k$th row with respect to all the $\ell$th rows, $\ell > k$.

Although this decomposition is not unique,
the $k$th row of $W(\theta)$ is the same as
the $k$th row of the partial product $G^{(k)}(\theta_k) \dots G^{(1)}(\theta_1)$.
As a result, let $X(\theta) = W(\theta)Z$,
we observe that the subset of angles in $\{\theta_{i, j}: 1 \leq i \leq k, i < j \leq d \} = \{\theta_i: 1 \leq i \leq k \}$
fully determines the $k$th element of $X$.
We define a support of $\theta$ as
\begin{equation}\label{support}
  \Theta = \left\{ \theta_{i, j}: \left\{
                               \begin{array}{ll}
                                 0 \leq \theta_{1, j} \leq 2 \pi, &  \\
                                 0 \leq \theta_{i, j} < \pi, & i \neq 1.
                               \end{array}
                              \right. \right\},
\end{equation}
and its subset with respect to $\theta_i$ as $\Theta_i$.
\citet{matteson2011dynamic} proved that there is a unique inverse mapping of $W \in \mathcal{SO}(d)$
into $\theta \in \Theta$, which is continuous if either all elements on the main-diagonal of $W$ are positive, or all
elements of $W$ are nonzero.

Unfortunately, the non-identification issue regarding $W$ and $X$ still exists
because the sign and order of the components are not identifiable.
Given any signed permutation matrix $P_\pm$, (\ref{ica2}) is equivalent to
\begin{equation*}
(P_\pm X) = P_\pm X = P_\pm WZ = (P_\pm W)Z,
\end{equation*}
where $P_\pm X$ and $P_\pm W$ become an alternative to $X$ and $W$, as the new ICs and unmixing matrix.
However, the identification up to a
signed permutation is adequate in terms of modeling multivariate data by linear combinations of ICs.
To make a fair comparison between different estimates, a metric invariant to the three ambiguities,
scale, sign, and order of the ICs will be presented in Section \ref{sim}.

Let $\mathbf{Y} \in \mathbb{R}^{n \times d}$ be an i.i.d.\ sample of observations from $F_Y$,
where $\mathbf{Y}_j \in \mathbb{R}^{n}$ is an i.i.d.\ sample of observations from $F_{Y_j}$, $j = 1, \dots, d$.
Let $\widehat{\Sigma}_\mathbf{Y}$ be the sample covariance matrix of $\mathbf{Y}$,
and $\widehat{H} = \widehat{\Sigma}_\mathbf{Y}^{-1/2}$ be the estimated uncorrelating matrix.
Although $\Sigma_Y$ is unknown in practice, the sample covariance is a consistent
estimate under the finite second-moment assumption,
i.e., $\widehat{\Sigma}_\mathbf{Y} \overset{a.s.}{\longrightarrow} \Sigma_Y$ as $n \rightarrow \infty$.
Let ${\widehat{\mathbf{Z}}} = \mathbf{Y} \widehat{H}' \in \mathbb{R}^{n \times d}$ be the estimated uncorrelated observations,
such that $\widehat{\Sigma}_{{\widehat{\mathbf{Z}}}} = I_d$,
and ${\Sigma}_{{\widehat{\mathbf{Z}}}} \overset{a.s.}{\longrightarrow} I_d$ as $n \rightarrow \infty$.

To simplify notation, we assume that ${\mathbf{Z}}$, an uncorrelated i.i.d.\ sample is given, with mean zero and unit variance.
Let $\mathbf{X}(\theta) = \mathbf{Z}W(\theta)' \in \mathbb{R}^{n \times d}$ be a sample of $X$.
Then we estimate $W(\theta)$ through $\theta$, and define an ICA estimator as
\begin{equation}\label{opt}
\widehat{\theta} = \underset{\theta \in \Theta}{\arg \min} f(\mathbf{X}(\theta)) = \underset{\theta \in \Theta}{\arg \min} f(\mathbf{Z}W(\theta)'),
\end{equation}
where $f$ is an objective function measuring the mutual dependence among $\mathbf{X}(\theta)$.
Given the estimate $\widehat{\theta}$, the estimated unmixing matrix is $\widehat{W} = W(\widehat{\theta})$,
and the estimated ICs are $\widehat{\mathbf{X}} = \mathbf{X}(\widehat{\theta}) = \mathbf{Z}\widehat{W}' = \mathbf{Z}W(\widehat{\theta})'$.


\section{APPLYING MDM TO ICA}\label{mdmica}

We reduce the estimation of ICs to the problem of choosing the function $f$ in (\ref{opt}), which is expected to
be a measure of mutual dependence.
Following \citet{matteson2017independent},
we primarily focus on distance-based energy statistics because of their compact representations
as expectations of pairwise Euclidean distances,
while all the results can be easily extended to kernel-based MMDs
according to the equivalence between these two classes in \citet{sejdinovic2013equivalence}.

We use $(\cdot, \cdot, \dots, \cdot)$ to concatenate (vector) components into a vector.
Let $t = (t_1, \dots, t_d), X = (X_1, \dots, X_d) \in \mathbb{R}^{p}$ where
$t_j, X_j \in \mathbb{R}^{p_j}$, $p_j$ is a marginal dimension, $j = 1, \dots, d$,
and $p = \sum_{j=1}^d p_j$ is the total dimension.
The subset of components to the right of $X_c$ is denoted by $X_{c^+} = (X_{c+1}, \dots, X_{d})$, $c = 0, 1, \dots, d-1$.
The subset of components excluding $X_c$ is denoted by $X_{-c} = ({X}_1, \dots, {X}_{c-1}, X_{c^+})$, $c = 1, \dots, d-1$.
The ``$X$'' under the assumption that $X_1, \dots, X_d$ are mutually independent
is denoted by $\widetilde{X} = (\widetilde{X}_1, \dots, \widetilde{X}_d)$,
where $\widetilde{X}_j \overset{\mathcal{D}}{=} X_j$, $j = 1, \dots, d$, $\widetilde{X}_1, \dots, \widetilde{X}_d$
are mutually independent, while $X, \widetilde{X}$ are independent.
Let $X', X''$ be independent copies of $X$ such that $X', X''$ have the same distribution as $X$, while they are all independent,
i.e., $X, X', X'' \overset{i.i.d.}{\sim} F_X$, and $\widetilde{X}'$ be an independent copy of $\widetilde{X}$.
The Euclidean norm of $X$ is denoted by $|X|$.
The weighted $\mathcal{L}_2$ norm $\|\cdot\|_w$ of any complex-valued function $\eta(t)$ is defined by $\|\eta(t)\|^2_w = \int_{\mathbb{R}^{p}} |\eta(t)|^2 w(t) \,dt$ where $|\eta(t)|^2 = \eta(t)\overline{\eta(t)}$, $\overline{\eta(t)}$ is the complex conjugate of $\eta(t)$, and $w(t)$ is any positive weight function for which the integral exists.

Let $\mathbf{X} = \{X^k = (X_1^k, \dots, X_d^k): k = 1, \dots, n \}$ be an i.i.d.\ sample from $F_X$, the joint distribution of $X$, and let $\mathbf{X}_j = \{X^k_j: k = 1, \dots, n \}$ be the corresponding i.i.d.\ sample from $F_{X_j}$, the marginal distribution of $X_j$, $j = 1, \dots, d$, such that $\mathbf{X} = \{\mathbf{X}_1, \dots, \mathbf{X}_d\}$.
Denote the joint characteristic function of $X$ as $\phi_{X}(t) = \textrm{E}[e^{i \langle t, X\rangle}]$
and its empirical version as $\phi^n_{X}(t) = \frac{1}{n} \sum_{k=1}^n e^{i \langle t, X^k\rangle}$,
and the joint characteristic function of $\widetilde{X}$ as
$\phi_{\widetilde{X}}(t) = \prod_{j = 1}^d \textrm{E}[e^{i \langle t_j, X_j\rangle}]$,
and its empirical version as
$\phi_{\widetilde{X}}^n(t) = \prod_{j = 1}^d (\frac{1}{n} \sum_{k=1}^n e^{i \langle t_j, X^k_j\rangle})$.
In addition, a simplified empirical version of $\phi_{\widetilde{X}}(t)$ is defined by
$\phi_{\widetilde{X}}^{n\star}(t) = \frac{1}{n}\sum_{k=1}^n e^{i \langle t, (X_1^k, \dots, X_d^{k+d-1}) \rangle}$ to substitute $\phi_{\widetilde{X}}^{n}(t)$
as a simplification, where $X_j^{n+k}$ is interpreted as $X_j^k$ for $k > 0$.

\subsection{DISTANCE COVARIANCE $(d = 2)$}

\citet{szekely2007measuring} proposed distance covariance to capture
non-linear and non-monotone pairwise dependence between two random vectors, i.e., $X = (X_1, X_2)$.

The nonnegative distance covariance $\mathcal{V}(X)$ is defined by
\begin{eqnarray*}
\mathcal{V}^2(X) &=& \|\phi_{X}(t) - \phi_{\widetilde{X}}(t)\|^2_{w_1} \\
&=& \int_{\mathbb{R}^{p}} |\phi_{X}(t) - \phi_{\widetilde{X}}(t)|^2 w_1(t) \,dt, \nonumber
\end{eqnarray*}
where the weight $w_1(t) = (K_{p_1} K_{p_2} |t_1|^{p_1+1} |t_2|^{p_2+1})^{-1}$, $K_q = \frac{2 \pi^{q/2} \Gamma(1/2) }{2 \Gamma((q + 1)/2)}$,
and $\Gamma$ is the gamma function.

An equivalence to pairwise independence is implied by the definition of $\mathcal{V}(X)$. If $\textrm{E}|X| < \infty$, then $\mathcal{V}(X) \in [0, \infty)$, and $\mathcal{V}(X) = 0$ if and only if $X_1, X_2$ are pairwise independent. In addition, if $\textrm{E}|X_1X_2| < \infty$, $\mathcal{V}^2(X)$ can be interpreted as expectations
\begin{eqnarray*}
\mathcal{V}^2(X) & = & \textrm{E}|X_1 - X_1'||X_2 - X_2'| \\
& & + \, \textrm{E}|X_1 - X_1'|\textrm{E}|X_2 - X_2'|  \nonumber \\
& & - \, 2\textrm{E}|X_1 - X_1'||X_2 - X_2''|. \nonumber
\end{eqnarray*}

We estimate $\mathcal{V}(X)$ by replacing the characteristic functions with the empirical characteristic functions from the sample.
The nonnegative empirical distance covariance $\mathcal{V}_n(\mathbf{X})$ is defined by
$\mathcal{V}_n^2(\mathbf{X}) = \|\phi^n_{X}(t) - \phi^n_{\widetilde{X}}(t)\|^2_{w_1} = \int_{\mathbb{R}^{p}} |\phi^n_{X}(t) - \phi^n_{\widetilde{X}}(t)|^2 w_1(t) \,dt$,
which can be interpreted as complete V-statistics
\begin{eqnarray*}
\mathcal{V}_n^2(\mathbf{X}) &=& \frac{1}{n^2}\sum_{k, \ell=1}^n |X_1^k - X_1^\ell||X_2^k - X_2^\ell| \\
&& \hspace{-1.5cm} + \, \frac{1}{n^2} \sum_{k,\ell=1}^n |X_1^k - X_1^\ell| \frac{1}{n^2} \sum_{k,\ell=1}^n |X_2^k - X_2^\ell| \\
&& \hspace{-1.5cm} - \, \frac{2}{n^3} \sum_{k,\ell,m = 1}^n |X_1^k - X_1^\ell||X_2^k - X_2^m|.
\end{eqnarray*}

Calculating $\mathcal{V}_n^2(\mathbf{X})$ via the symmetry of Euclidian distances has the time complexity $O(n^2)$ .
If $\textrm{E}|X| < \infty$, then we have $\mathcal{V}_n(\mathbf{X}) \overset{a.s.}{\longrightarrow} \mathcal{V}(X)$ as $n \rightarrow \infty$.

\citet{jin2017generalizing} generalized distance covariance to three mutual dependence measures
capturing any form of mutual dependence between multiple random vectors,
which include the asymmetric, symmetric, and complete measures below.

\subsection{ASYMMETRIC AND SYMMETRIC MEASURES $(d \geq 2)$}

The asymmetric and symmetric measures of mutual dependence $\mathcal{R}(X), \mathcal{S}(X)$ are defined by
\begin{eqnarray}\label{asym_measure}
\mathcal{R}(X) &=& \sum_{c=1}^{d-1} \mathcal{V}^2((X_c, X_{c^+})), \\
\mathcal{S}(X) &=& \sum_{c=1}^d \mathcal{V}^2((X_c, X_{-c})).
\end{eqnarray}

Analogous to $\mathcal{V}(X)$,
if $\textrm{E}|X| < \infty$, then $\mathcal{R}(X), \mathcal{S}(X) \in [0, \infty)$,
and $\mathcal{R}(X), \mathcal{S}(X) = 0$ if and only if $X_1, \dots, X_d$ are mutually independent.

Correspondingly, the empirical asymmetric and symmetric measures of mutual dependence
$\mathcal{R}_n(\mathbf{X}), \mathcal{S}_n(\mathbf{X})$ are defined by
$\mathcal{R}_n(\mathbf{X}) = \sum_{c=1}^{d-1}\mathcal{V}_n^2((\mathbf{X}_c, \mathbf{X}_{c^+}))$,
$\mathcal{S}_n(\mathbf{X}) = \sum_{c=1}^d \mathcal{V}_n^2((\mathbf{X}_c, \mathbf{X}_{-c}))$,
which can be implemented with the time complexity $O(n^2)$.
If $\textrm{E}|X| < \infty$, then we have $\mathcal{R}_n(\mathbf{X}) \overset{a.s.}{\longrightarrow} \mathcal{R}(X)$ and
$\mathcal{S}_n(\mathbf{X}) \overset{a.s.}{\longrightarrow} \mathcal{S}(X)$ as $n \rightarrow \infty$.

\subsection{COMPLETE MEASURE $(d \geq 2)$}

The complete measure of mutual dependence $\mathcal{Q}(X)$ is defined by
\begin{eqnarray*}
  \mathcal{Q}(X) &=& \|\phi_{X}(t) - \phi_{\widetilde{X}}(t)\|^2_{w_2} \\
   &=& \int_{\mathbb{R}^{p}} |\phi_{X}(t) - \phi_{\widetilde{X}}(t)|^2 w_2(t) \,dt, \nonumber
\end{eqnarray*}
where $w_2(t) = (K_p |t|^{p+1})^{-1}$, $K_q = \frac{2 \pi^{q/2} \Gamma(1/2) }{2 \Gamma((q + 1)/2)}$, and $\Gamma$ is the gamma function.

An equivalence to mutual independence is implied by the definition of $\mathcal{Q}(X)$ as well.
If $\textrm{E}|X| < \infty$, then $\mathcal{Q}(X) \in [0, \infty)$, and $\mathcal{Q}(X) = 0$ if and only if $X_1, \dots, X_d$ are mutually independent. In addition, $\mathcal{Q}(X)$ can be interpreted as expectations
\begin{equation*}
\mathcal{Q}(X) = \textrm{E}|X - \widetilde{X}'| + \textrm{E}|X' - \widetilde{X}| - \, \textrm{E}|X - X'| - \textrm{E}|\widetilde{X} - \widetilde{X}'|.
\end{equation*}

We estimate $\mathcal{Q}(X)$ by two empirical versions.
One is the empirical complete measure of mutual dependence $\mathcal{Q}_n(\mathbf{X})$, defined by
$\mathcal{Q}_n(\mathbf{X}) = \|\phi^n_{X}(t) - \phi^n_{\widetilde{X}}(t)\|^2_{w_2} = \int_{\mathbb{R}^{p}} |\phi^n_{X}(t) - \phi^n_{\widetilde{X}}(t)|^2 w_2(t) \,dt$,
which can be interpreted as complete V-statistics.
We skip the details of $\mathcal{Q}_n$ and will not apply it to ICA, since it is computationally prohibitive with the time complexity $O(n^{2d})$.
Another one is the simplified empirical complete measure of mutual dependence $\mathcal{Q}_n^\star(\mathbf{X})$, defined by
$\mathcal{Q}_n^\star(\mathbf{X}) = \|\phi_{X}^n(t) - \phi_{\widetilde{X}}^{n\star}(t) \|^2_{w_2} = \int_{\mathbb{R}^{p}} |\phi^n_{X}(t) - \phi^{n\star}_{\widetilde{X}}(t)|^2 w_2(t) \,dt$, which can be interpreted as incomplete V-statistics
\begin{eqnarray*}
\mathcal{Q}_n^\star(\mathbf{X}) &=& \frac{2}{n^{2}}\sum_{k, \ell = 1}^n |X^k - (X_1^{\ell}, \dots, X_d^{\ell + d - 1})| \\
&& \hspace{-1.5cm} + \, \frac{1}{n^2} \sum_{k,\ell=1}^n |X^k - X^\ell| \\
&& \hspace{-1.5cm} - \, \frac{1}{n^{2}} \sum_{k, \ell = 1}^n  |(X_1^{k}, \dots, X_d^{k + d - 1}) - (X_1^{\ell}, \dots, X_d^{\ell + d - 1})|.
\end{eqnarray*}

The naive implementation of $\mathcal{Q}_n^\star(\mathbf{X})$ has the time complexity $O(n^2)$.
If $\textrm{E}|X| < \infty$, then
$\mathcal{Q}_n(\mathbf{X}), \mathcal{Q}^\star_n(\mathbf{X}) \overset{a.s.}{\longrightarrow} \mathcal{Q}(X)$ as $n \rightarrow \infty$.

\subsection{MDMICA APPROACH AND ITS ASYMPTOTIC PROPERTIES}

Inspired by the nice statistical properties of MDMs, we propose an ICA approach, MDMICA based on MDMs.
To be specific, we define three MDMICA estimators, i.e., MDMICA (asy), MDMICA (sym), and MDMICA (com)
by applying $f(\mathbf{X}) = \mathcal{R}_n(\mathbf{X}), \mathcal{S}_n(\mathbf{X}), \mathcal{Q}_n^\star(\mathbf{X})$
in (\ref{opt}) respectively as
\begin{equation*}\label{3est}
\widehat{\theta}^{\textrm{asy}}_n = \min_{\theta \in \Theta} \mathcal{R}_n(\mathbf{X}(\theta)) = \min_{\theta \in \Theta} \mathcal{R}_n(\mathbf{Z}W(\theta)'),
\end{equation*}
and similar expressions follow for $\widehat{\theta}^{\textrm{sym}}_n, \widehat{\theta}^{\textrm{com}}_n$.
Further, we define another estimator, MDMICA (hsic), by applying dHSIC in the same way.

Since the ICA model only allows scalar components,
we apply a special case of MDM to ICA
where the marginal dimension $p_j = 1$, $j = 1, \dots, d$, and the total dimension $p = d$.
Without loss of generality,
we assume that E$(Y) = 0$ and Cov$(Y) = I_d$,
and therefore $Z = Y$ and $\mathbf{Z} = \mathbf{Y}$ throughout this section.
Let $\overline{\Theta}$ denote a large enough compact subset of the space $\Theta$ defined by (\ref{support}).
The asymptotic properties of the MDMICA estimators are derived as follows.

\begin{theorem}\label{est-thm-1}
If $Y$ has a nonsingular, continuous distribution $F_Y$ with E$|Y|^2 < \infty$,
if there exists a unique minimizer $\theta_0 \in \overline{\Theta}$ of
(\ref{asym_measure}), and if $W({\theta_0})$ satisfies the conditions for a unique continuous inverse to exist,
then $\widehat{\theta}^{\textrm{asy}}_n \overset{a.s.}{\longrightarrow} \theta_0$ as $n \rightarrow \infty$.
\end{theorem}
When the ICA model is misspecified, convergence to the pseudo-true
value $\theta_0$ is obtained. Under similar conditions,
$\widehat{\theta}^{\textrm{sym}}_n, \widehat{\theta}^{\textrm{com}}_n$ also converges a.s.\
as $n \rightarrow \infty$ due to similar arguments.

We then establish the root-$n$ consistency of the MDMICA estimators
under some regularity conditions no matter whether the ICA model holds or it is misspecified.

\begin{theorem}\label{est-thm-2}
If the assumptions of Theorem \ref{est-thm-1} hold, and if the ICA model assumptions hold,
then $|\widehat{\theta}^{\textrm{asy}}_n - \theta_0| = O_P(n^{-1/2})$.
\end{theorem}

\begin{theorem}\label{est-thm-3}
If the ICA model is misspecified but the remaining assumptions stated in Theorem \ref{est-thm-2}
hold, and if E$[\frac{\partial}{\partial \theta} \mathcal{R}_n(\mathbf{X}(\theta)) |_{\theta = \theta_0}] = o_P(n^{-1/2})$,
where $\theta_0$ denotes the pseudo-true value, then $|\widehat{\theta}^{\textrm{asy}}_n - \theta_0| = O_P(n^{-1/2})$.
\end{theorem}
Under similar conditions, $\widehat{\theta}^{\textrm{sym}}_n, \widehat{\theta}^{\textrm{com}}_n$ are also consistent
as $n \rightarrow \infty$ due to similar arguments.

The proofs of Theorem \ref{est-thm-1}, \ref{est-thm-2}, and \ref{est-thm-3}
are similar to those of Theorem 2.1, 2.2, and Corollary 2.1 in \citet{matteson2017independent} respectively,
considering the same nature of $\mathcal{R}_n(\mathbf{X}), \mathcal{S}_n(\mathbf{X}), \mathcal{Q}_n^\star(\mathbf{X})$ as energy statistics,
and replacing the empirical cumulative distribution function (ECDF) with the identity function in derivations.

\section{IMPROVING INITIALIZATION OF LOCAL METHODS}\label{global}

In the literature, there are two primary schemes to estimate ICs with regard to how the optimization is implemented.
For one, the components are extracted one at a time, known as the deflation scheme.
For another, the components are extracted simultaneously, known as the parallel scheme.
The deflation scheme has the advantage of lower computational cost over the parallel scheme.
While the parallel scheme enjoys greater statistical efficiency,
since the deflation scheme accumulates estimation uncertainty at each step in its sequential procedure.

For our ICA framework,
the objective function $f$ in (\ref{opt}) has $d(d-1)/2$ parameters $\theta_{i, j} \in \theta$,
which can be estimated in both deflation (sequential) and parallel (joint) manners.
Specifically, the deflation scheme estimates all $\theta_{i, j} \in \theta$ for each $i$ at a time,
while the parallel scheme estimates all $\theta_{i, j} \in \theta$ together at once.

In view of the special structures of associated measures,
both deflation and parallel schemes are appropriate for MDMICA (asy),
denoted by MDMICA (asy, def) and MDMICA (asy, par),
while MDMICA (sym), MDMICA (com), and MDMICA (hsic) only fit the parallel scheme.
The MDMICA algorithms for both deflation and parallel schemes are described in Alg. \ref{alg1} below.
\begin{algorithm}
\caption{MDMICA ($\mathbf{Z}, f$)}\label{alg1}
\begin{algorithmic}
\STATE 1. Initialize $\theta$ and $W(\theta)$ via $\theta$.
\STATE 2. (deflation scheme)
\STATE \quad \textbf{for} $i = 1, \cdots, d-1$ \textbf{do}
\STATE \quad \quad a. Solve $\widehat{\theta}_i = \underset{\theta_i \in \Theta_i}{\arg \min} f(\mathbf{Z}W(\theta)')$ using newton-\\
\quad \quad type local optimization.
\STATE \quad \quad b. Update $\theta_i \leftarrow \widehat{\theta}_i$.
\STATE \quad \textbf{end for}

\STATE 2'. (parallel scheme)
\STATE \quad \parbox[t]{\dimexpr\linewidth-\algorithmicindent}{Solve $\widehat{\theta} = \underset{\theta \in \Theta}{\arg \min} f(\mathbf{Z}W(\theta)')$ using newton-type local optimization.}

\STATE 3. \parbox[t]{\dimexpr\linewidth-\algorithmicindent}{Output $\widehat{\theta} = \{\widehat{\theta}_i: 1 \leq i \leq d - 1\}$,
$\widehat{W} = W(\widehat{\theta})$, and $\widehat{\mathbf{X}} = \mathbf{Z}W(\widehat{\theta})'$.}

\end{algorithmic}
\end{algorithm}

Estimating $\theta$ through (\ref{opt}) involves minimization of a
non-convex but locally convex objective function $f$, which requires initialization and iterative algorithms.
The default method for MDMICA is a Newton-type local
optimization method, for which we explore two ways of finding a good initialization.

The first way is to perform a random sampling method,
Latin hypercube sampling (LHS) \citep{mckay2000comparison} uniformly over the space $\Theta$ to obtain a number of parameter values.
Then we evaluate the objective function at each value and record the value minimizing it,
which is used to initialize the subsequent local optimization algorithm.
Based on our experience, the number of parameter values sampled should grow with the dimension.

The second way is to take advantage of a global optimization method,
Bayesian optimization (BO) \citep{mockus1994application}, where the
objective function $f$ is treated as a black box.
It is applicable when the function is expensive to evaluate, the derivative is unavailable,
or the optimization problem is non-convex.
Bayesian optimization is one of the most efficient approaches
in terms of the number of function evaluations consumed,
as \citet{jones2001taxonomy}, \citet{brochu2010tutorial}, \citet{snoek2012practical}
illustrated that it outperforms other state-of-the-art global optimization
algorithms on a number of challenging problems.


Bayesian optimization models the objective with respect to the parameter values
as a Gaussian process,
for which we adopt two popular kernels, squared exponential (exp) and Mat\'{e}rn 5/2 (Mat\'{e}rn).
A prior is set over the objective function and then
updated with actual evaluations to get a posterior using the Bayesian technique.
The utility-based selection of the next evaluation point on the objective
function trades off between exploration and exploitation.

\section{SIMULATION STUDIES}\label{sim}

In this section, we evaluate the performance of our MDMICA estimators
by performing simulations similar to \citet{matteson2017independent},
and compare them with the FastICA estimator,
the Infomax estimator, and the JADE estimator.
MDMICA (asy) is omitted because it is the same as dCovICA.
Moreover, we elaborate on the implementation and error metric of ICA.

Furthermore, we try various options for each estimator.
For FastICA, we evaluate three functions used to approximate negentropy in both deflation and parallel schemes,
logarithm of hyperbolic cosine (logcosh), kurtosis (kur), and exponential (exp).
For Infomax, we evaluate three nonlinear (squashing) functions, hyperbolic
tangent (tanh), logistic (log), and extended Infomax (ext).
For MDMICA (hsic), we investigate the Gaussian (gau) kernel.
However, FastICA (kur) and FastICA (exp) are omitted
since their performance is quite similar to that of FastICA (logcosh).
Similarly, Infomax (log) and Infomax (ext) are omitted.

\begin{figure}[!ht]
\begin{center}
\centerline{\includegraphics[width=\columnwidth]{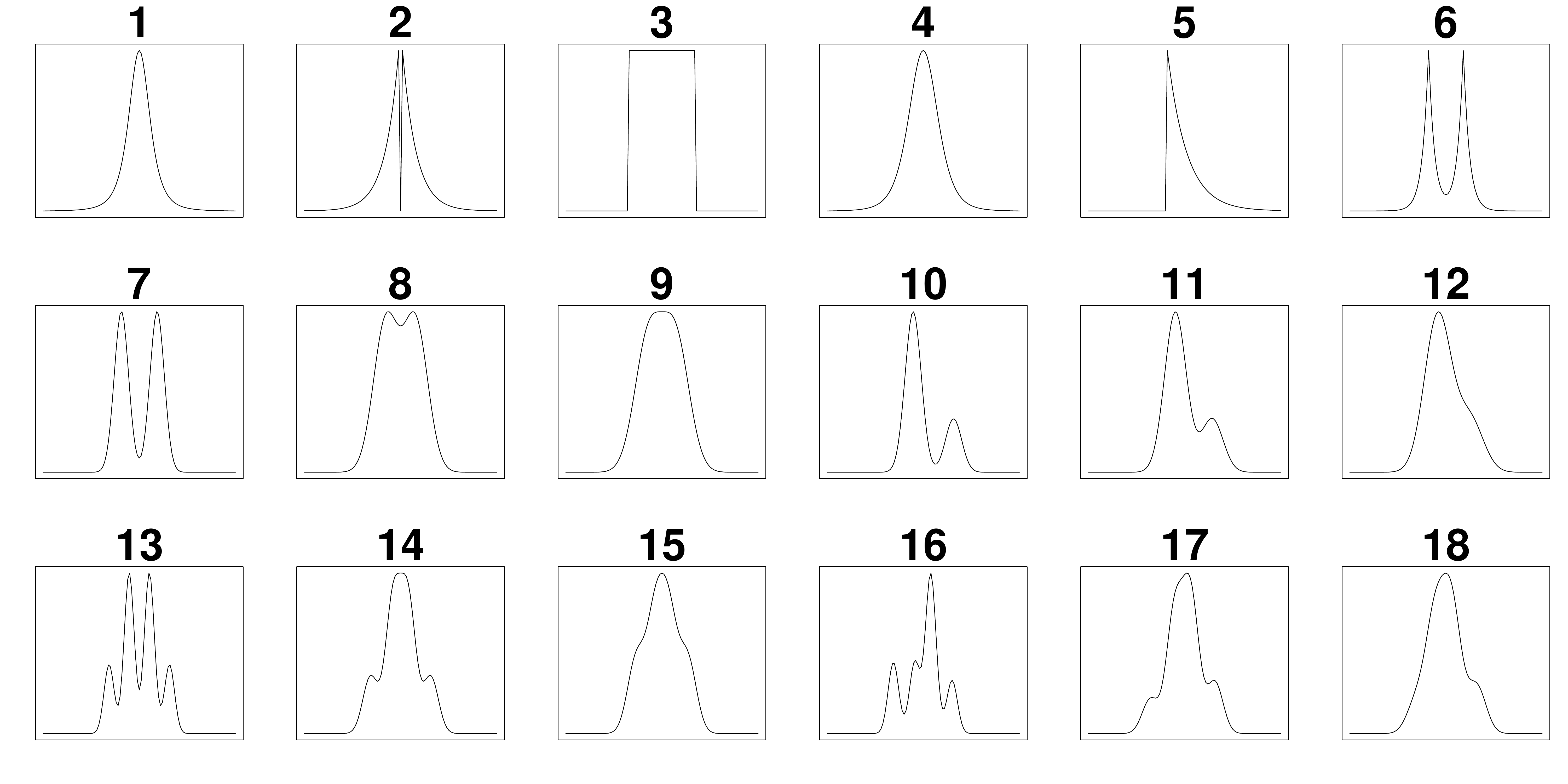}}
\caption{Density plots of the 18 distributions.}\label{fig0}
\end{center}
\vskip -0.2in
\end{figure}

We simulate the ICs $\mathbf{X} \in \mathbb{R}^{n \times d}$ from eighteen distributions using \texttt{rjordan}
in the R package \texttt{ProDenICA} \citep{hastie2010prodenica} with sample size $n$ and dimension $d$.
See Figure \ref{fig0} for the density functions of the eighteen distributions.
Then we generate a mixing matrix $M \in \mathbb{R}^{d\times d}$ with condition number
between 1 and 2 using \texttt{mixmat} in the R package \texttt{ProDenICA} \citep{hastie2010prodenica},
and obtain the observations $\mathbf{Y} = \mathbf{X}M'$,
which are centered by their sample mean and then prewhitened by their sample covariance to obtain uncorrelated observations
$\mathbf{Z} = \mathbf{Y}\widehat{H}'$. Finally, we obtain the estimate $\widehat{W}$ based on $\mathbf{Z}$ via (\ref{opt}),
and evaluate the estimation accuracy by comparing the estimate $\widehat{W}$ to the ground truth $W_0 = (\widehat{H}M)^{-1}$.
Moreover, the Newton-type local optimization is implemented by
\texttt{nlm} in the R package \texttt{stats} \citep{r2014},
and Bayesian optimization is implemented by \texttt{BayesianOptimization}
in the R package \texttt{rBayesianOptimization} \citep{yan2016bayes}.

To take the uncertainty in both prewhitening the observations and estimating the ICs
into account when comparing the estimates from different approaches,
we use the metric MD proposed by \citet{ilmonen2010new}
to measure the error between an estimate $\widehat{W}$ and the corresponding truth $W_0$, which is defined as
\begin{equation*}
\textrm{MD}(\widehat{M}, M) = \frac{1}{\sqrt{d-1}} \inf_{P, D} \|P D \widehat{W} W_0^{-1} - I_d \|_F,
\end{equation*}
where $\|\cdot\|_F$ denotes the Frobenius norm,
$P$ is a $d \times d$ permutation matrix,
and $D$ is a $d \times d$ diagonal matrix with nonzero diagonal elements.
MD is invariant to the three ambiguities associated with ICA as a result of taking the infimum,
for which the optimal $P, D$ are solved by the Hungarian method \citep{papadimitriou1998combinatorial}.


\begin{experiment}\label{exp1}\textit{[Different distributions of ICs]}
{\rm
We sample $\mathbf{X}$ from one distribution in the eighteen distributions, with $d = 3$, $n = 1000$.
We obtain $10d$ points using LHS, and select the best initial point.
See Figure \ref{fig1} for the error metrics of all eighteen distributions with 100 trials.

\begin{figure*}[!ht]
\begin{center}
\centerline{\includegraphics[width=2\columnwidth]{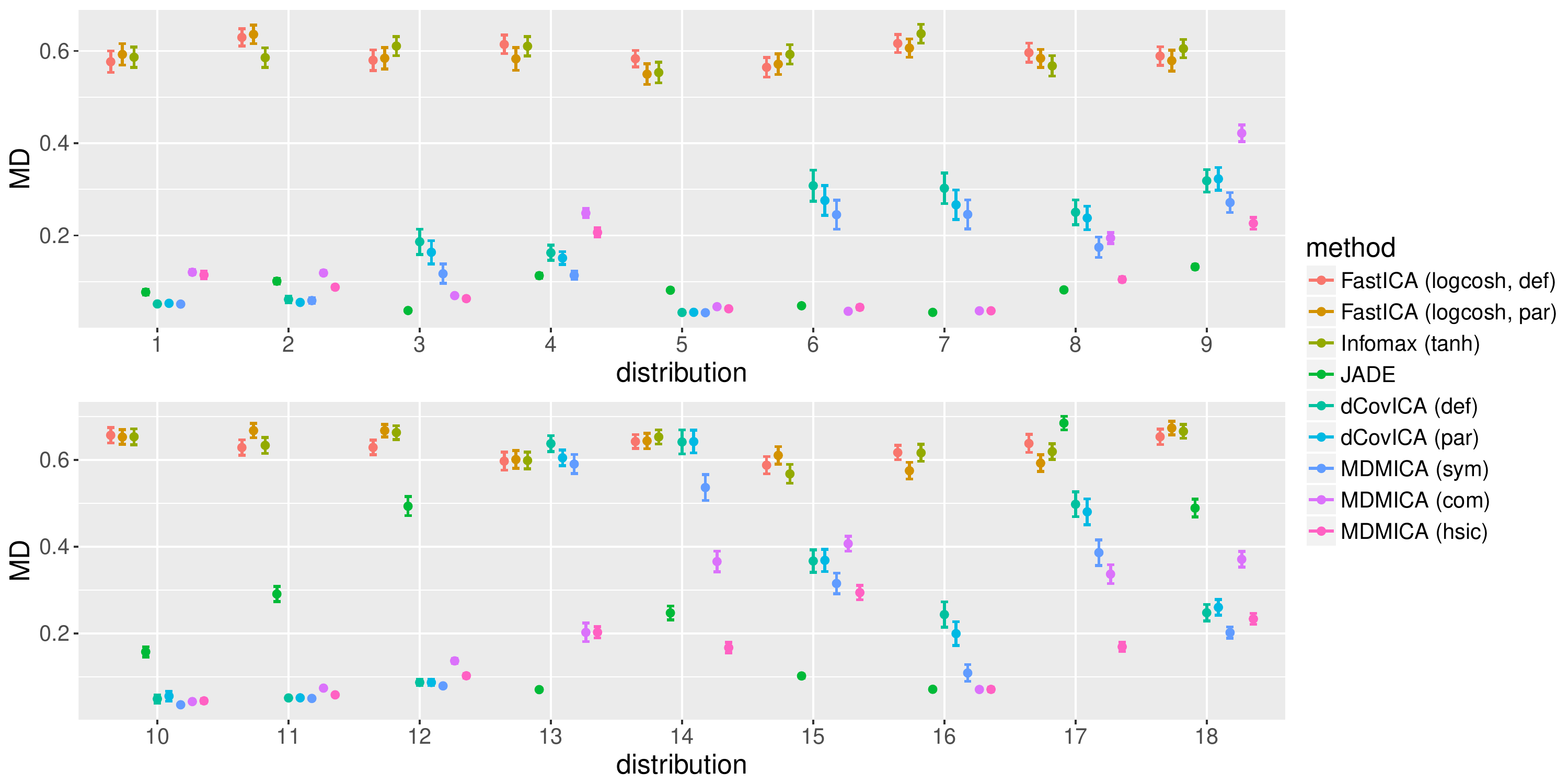}}
\caption{Error metrics (mean $\pm$ standard error) of all eighteen distributions with 100 trials for Model \ref{exp1}.}\label{fig1}
\end{center}
\end{figure*}

MDMICA achieves competitive results with JADE and dCovICA,
and also outperforms FastICA and Infomax in most cases.
MDMICA (sym) is equal and often better than dCovICA,
while they have similar performance due to their similar structures.
MDMICA (hsic) is equal and often better than MDMICA (com),
while they have similar performance due to their similar structures.
Further, MDMICA (com) and MDMICA (hsic) are less sensitive to different distributions
than dCovICA and MDMICA (sym) in general.
Lastly, there is no remarkable difference between the deflation and parallel schemes.




}

\end{experiment}

\begin{experiment}\label{exp2}\textit{[Different dimensions of ICs]}
{\rm
We sample $\mathbf{X}$ from one distribution in the eighteen distributions, with $d \in \{2, 3, 4\}$, $n = 1000$.
We pick $10d$ points using LHS, and select the best initial point.
See Figure \ref{fig2} for the error metrics of the 1st distribution with 100 trials.

\begin{figure}[!ht]
\begin{center}
\centerline{\includegraphics[width=\columnwidth]{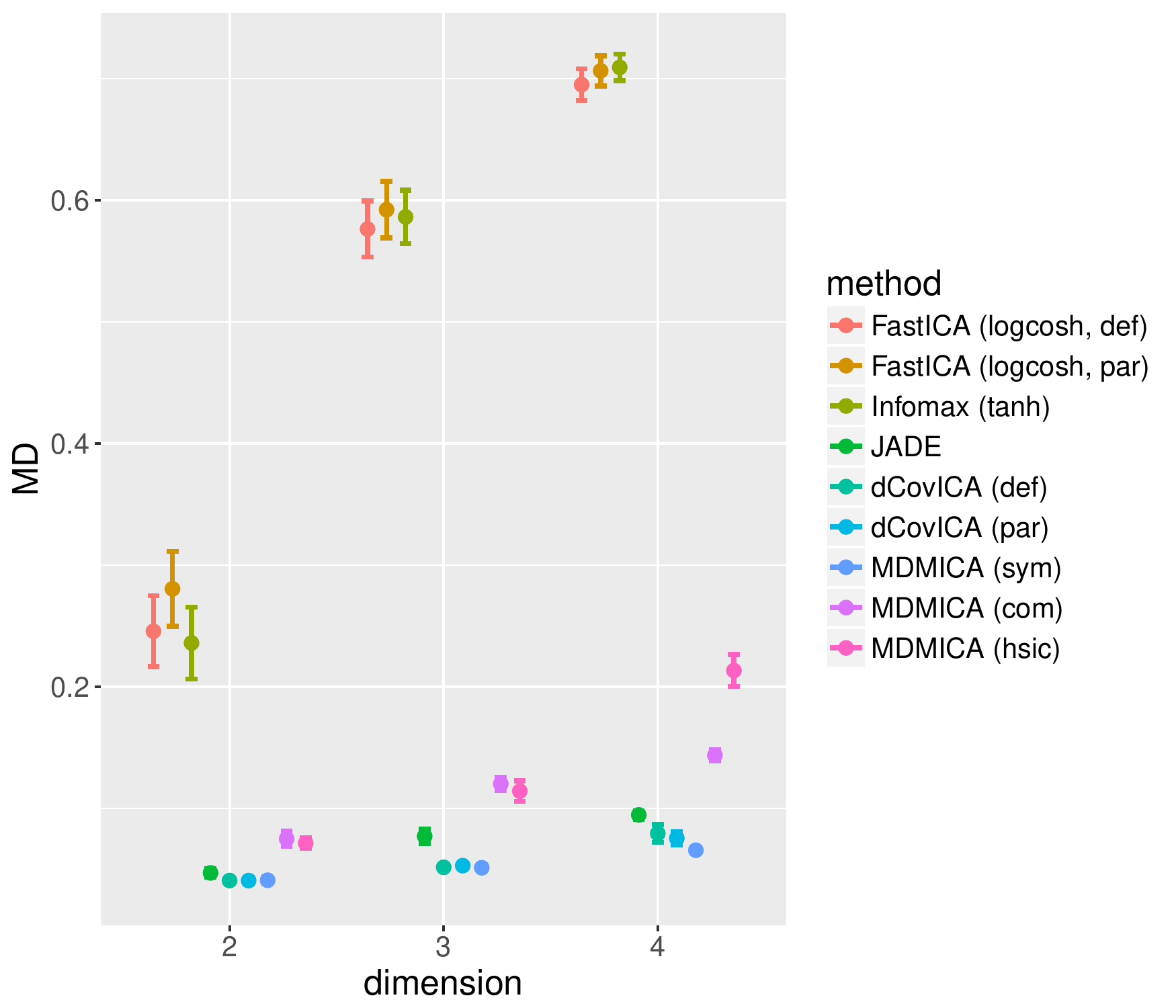}}
\caption{Error metrics (mean $\pm$ standard error) of the 1st distribution with 100 trials for Model \ref{exp2}.}\label{fig2}
\end{center}
\vspace{-0.5cm}
\end{figure}


The errors of all estimators increase as the dimension $d$ grows.
As in the previous model, JADE, dCovICA, and MDMICA have similar performance,
and significantly outperform FastICA and Infomax.



}

\end{experiment}

\begin{experiment}\label{exp3}\textit{[Different initializations of local optimization]}
{\rm
We sample $\mathbf{X}$ from $d$ randomly selected distributions of the eighteen distributions, with $d = 4$, $n = 1000$.
We implement three ways to select the initial point for the
Newton-type local optimization method. The first way is to sample one point using LHS, and then proceed.
The second way is to sample $10d$ points using LHS, and then select the
point out of $10d$ with the lowest objective.
The third way is to run $10d$ iterations using BO, with its initial points from $10d$ sampled points using LHS,
and then select the point out of $20d$ with the lowest objective.
See Table \ref{tab1} for the error metrics, objective values, and computational times
of the tuple as the (4th, 11th, 12th, 18th) distributions with 50 trials.

\begin{table*}[!ht]
\caption{Error metrics (mean $\pm$ standard error), objective values (mean $\pm$ standard error), and computational times (mean)
of the tuple as the (4th, 11th, 12th, 18th) distributions with 50 trials for Model \ref{exp3}.}
\label{tab1}
\begin{center}
\begin{small}
\begin{tabular}{c|c|c|c|r}
\hline
Estimator & Initialization & MD ($10^{-1}$) & Objective ($10^{-3}$) & Time (second) \\ \hline
FastICA (logcosh, def) & LHS ($10d$)                         & 7.157 $\pm$ 0.130 & - & 0.28 \\ \hline
FastICA (logcosh, par) & LHS ($10d$)                         & 6.902 $\pm$ 0.159 & - & 0.10 \\ \hline
Infomax (tanh)         & LHS (1)                             & 6.802 $\pm$ 0.166 & - & 0.07 \\ \hline
JADE                   & LHS (1)                             & 3.933 $\pm$ 0.215 & - & 0.01 \\ \hline

\multirow{4}{*}{dCovICA (def)}
                       & LHS (1)                             & 1.352 $\pm$ 0.144 & 3.997 $\pm$ 0.099& 34.83 \\
                       & LHS ($10d$)                         & 1.276 $\pm$ 0.119 & 3.988 $\pm$ 0.084 & 46.89 \\
                       & LHS ($10d$) + BO (exp)              & 1.246 $\pm$ 0.120 & 3.968 $\pm$ 0.080 & 2453.89 \\
                       & LHS ($10d$) + BO (Mat\'{e}rn)       & 1.250 $\pm$ 0.119 & 3.965 $\pm$ 0.084 & 8080.55 \\ \hline

\multirow{4}{*}{dCovICA (par)}
                       & LHS (1)                             & 1.359 $\pm$ 0.133 & 3.993 $\pm$ 0.089 & 189.75 \\
                       & LHS ($10d$)                         & 1.230 $\pm$ 0.093 & 3.891 $\pm$ 0.053 & 175.90 \\
                       & LHS ($10d$) + BO (exp)              & 1.183 $\pm$ 0.091 & 3.878 $\pm$ 0.054 & 2990.86 \\
                       & LHS ($10d$) + BO (Mat\'{e}rn)       & 1.105 $\pm$ 0.053 & 3.863 $\pm$ 0.056 & 8830.85 \\ \hline

\multirow{4}{*}{\textbf{MDMICA (sym)}}
                       & \textbf{LHS (1)}                             & \textbf{1.160 $\pm$ 0.104} & \textbf{7.057 $\pm$ 0.189} & \textbf{217.49} \\
                       & \textbf{LHS ($10d$)}                         & \textbf{1.099 $\pm$ 0.086} & \textbf{6.908 $\pm$ 0.095} & \textbf{195.86} \\
                       & \textbf{LHS ($10d$) + BO (exp)}              & \textbf{1.096 $\pm$ 0.086} & \textbf{6.908 $\pm$ 0.095} & \textbf{2959.62} \\
                       & \textbf{LHS ($10d$) + BO (Mat\'{e}rn)}       & \textbf{1.097 $\pm$ 0.086} & \textbf{6.908 $\pm$ 0.095} & \textbf{9063.81} \\ \hline

\multirow{4}{*}{\textbf{MDMICA (com)}}
                       & \textbf{LHS (1)}                             & \textbf{2.758} $\pm$ \textbf{0.267} & \textbf{2.010 $\pm$ 0.118} & \textbf{63.62} \\
                       & \textbf{LHS ($10d$)}                         & \textbf{2.194} $\pm$ \textbf{0.150} & \textbf{1.656 $\pm$ 0.019} & \textbf{32.97} \\
                       & \textbf{LHS ($10d$) + BO (exp)}              & \textbf{1.951} $\pm$ \textbf{0.123} & \textbf{1.651 $\pm$ 0.018} & \textbf{2892.11} \\
                       & \textbf{LHS ($10d$) + BO (Mat\'{e}rn)}       & \textbf{1.954} $\pm$ \textbf{0.123} & \textbf{1.642 $\pm$ 0.019} & \textbf{8529.42} \\ \hline

\multirow{4}{*}{\textbf{MDMICA (hsic)}}
                       & \textbf{LHS (1)}                             & \textbf{4.238 $\pm$ 0.319} & \textbf{1.319 $\pm$ 0.117} & \textbf{270.97} \\
                       & \textbf{LHS ($10d$)}                         & \textbf{2.674 $\pm$ 0.217} & \textbf{0.848 $\pm$ 0.030} & \textbf{301.82} \\
                       & \textbf{LHS ($10d$) + BO (exp)}              & \textbf{2.153 $\pm$ 0.169} & \textbf{0.778 $\pm$ 0.017} & \textbf{2774.90} \\
                       & \textbf{LHS ($10d$) + BO (Mat\'{e}rn)}       & \textbf{2.177 $\pm$ 0.198} & \textbf{0.789 $\pm$ 0.023} & \textbf{8437.69} \\ \hline

\hline
\end{tabular}
\end{small}
\end{center}
\end{table*}

The performance of dCovICA and MDMICA is greatly improved by selecting the best point from multiple initial points,
as LHS and LHS + BO produce smaller objective values and more accurate estimates than a single point with lower mean and standard error.
The reason is two-fold.
First, LHS and BO offer the subsequent local optimization method better initial points
in terms of lower objective, which leads to a better estimate in terms of lower objective as well.
Second, a better estimate with lower objective is likely to be a better
solution with lower MD, since the objective is a truly mutual dependence measure.
Moreover, LHS + BO has noticeable advantage over LHS alone for MDMICA (com) and MDMICA (hsic),
but only marginal advantage over LHS alone for dCovICA (def), dCovICA (par), and MDMICA (sym).

dCovICA and MDMICA take remarkably longer computational time than the others,
which makes sense because the optimization problem of dCovICA and MDMICA is especially difficult to solve,
as it has $d(d - 1)/2$ parameters and becomes high-dimensional quickly.
This obstacle in turn motivates us to improve the local optimization by choosing a better initialization point.

}

\end{experiment}

\begin{experiment}\label{exp5}\textit{[Misspecified ICA model]}
{\rm
We sample $\mathbf{X} = (\mathbf{X_1}, \mathbf{X_2})$ from one distribution in the eighteen distributions, with $n = 1000$.
Let $\mathbf{Y_1} = \mathbf{X_1}$, $\mathbf{Y_2} = (\mathbf{X_2})^2$.
We pick $10d$ points using LHS, and select the best initial point.
See Table \ref{tab3} for the results of the 1st distribution with 1 trial.

\begin{table*}[!ht]
\caption{Mutual dependence measures of observed components (before optimization, $\mathbf{Z}$) and
estimated independent components (after optimization, $\widehat{\mathbf{X}}$) with 1 trial for Model \ref{exp5}
(misspecified ICA model).}
\label{tab3}
\begin{center}
\begin{small}
\begin{tabular}{c|c|c|c|c|c|c}
\hline
Estimator & $\mathcal{R}_n(\mathbf{Z})$ ($10^{-3}$) & $\mathcal{R}_n(\widehat{\mathbf{X}})$
& $\mathcal{S}_n(\mathbf{Z})$ ($10^{-3}$) & $\mathcal{S}_n(\widehat{\mathbf{X}})$
& $\mathcal{Q}_n^\star(\mathbf{Z})$ ($10^{-4}$) & $\mathcal{Q}_n^\star(\widehat{\mathbf{X}})$ \\
\hline
FastICA (logcosh, def) & \multirow{9}{*}{0.548} & 0.531 & \multirow{9}{*}{1.097} & 1.062 & \multirow{9}{*}{2.797} & 3.088 \\
FastICA (logcosh, par) &            & 0.588 &            & 1.176 &             & 2.786 \\
Infomax (tanh)         &            & 0.606 &            & 1.212 &             & 3.081 \\
JADE                   &            & 1.031 &            & 2.062 &             & 3.330 \\
dCovICA (def)      &            & 0.441 &            & 0.882 &             & 2.677 \\
dCovICA (par)      &            & 0.441 &            & 0.882 &             & 2.677 \\
\textbf{MDMICA (sym)}     &            & \textbf{0.441} &            & \textbf{0.882} &             & \textbf{2.677} \\
\textbf{MDMICA (com)}     &            & \textbf{0.446} &            & \textbf{0.892} &             & \textbf{2.672} \\
\textbf{MDMICA (hsic)}     &            & \textbf{0.443} &            & \textbf{0.887} &             & \textbf{2.687} \\
\hline
\end{tabular}
\end{small}
\end{center}
\end{table*}

We use $\mathcal{R}_n, \mathcal{S}_n, \mathcal{Q}_n^\star$
to measure the mutual dependence between the components before (w.r.t.\ $\mathbf{Z}$) and after (w.r.t.\ $\widehat{\mathbf{X}}$) the optimization.
dCovICA and MDMICA successfully decreases the mutual dependence between the components through optimization,
while FastICA, Infomax, and JADE are unable to and even increase it.
Therefore, ICA methods based on mutual dependence measures outperform others in reducing the mutual dependence given that the ICA model is misspecified.
}
\end{experiment}

\section{IMAGE DATA}\label{data}

Fulfilling the task of unmixing vectorized images similar to \citet{virta2016projection},
we consider the three gray-scale images in the R package \texttt{ICS} \citep{nordhausen2008tools},
depicting a cat, a forest road, and a sheep respectively.
Each image is represented by a $130 \times 130$ matrix,
where each element indicates the intensity value of a pixel.
We standardize the three images such that the intensity values across all the pixels in each image have mean zero and unit variance.
Then we vectorize each image into a vector of length $130^2$,
and combine the vectors from all three images as a matrix $\mathbf{X}$,
with $d = 3$, $n = 130^2$.

We use \texttt{mixmat} in the R package \texttt{ProDenICA} \citep{hastie2010prodenica} again
to generate a mixing matrix $A \in \mathbb{R}^{p\times p}$,
and mix the three images to obtain the observations $\mathbf{Y} = \mathbf{X}A^T$,
which are centered by their sample mean, then prewhitened by their sample covariance to obtain uncorrelated observations
$\mathbf{Z} = \mathbf{Y}\widehat{H}^T$.

\begin{figure}[!ht]
\begin{center}
\centerline{\includegraphics[width=\columnwidth]{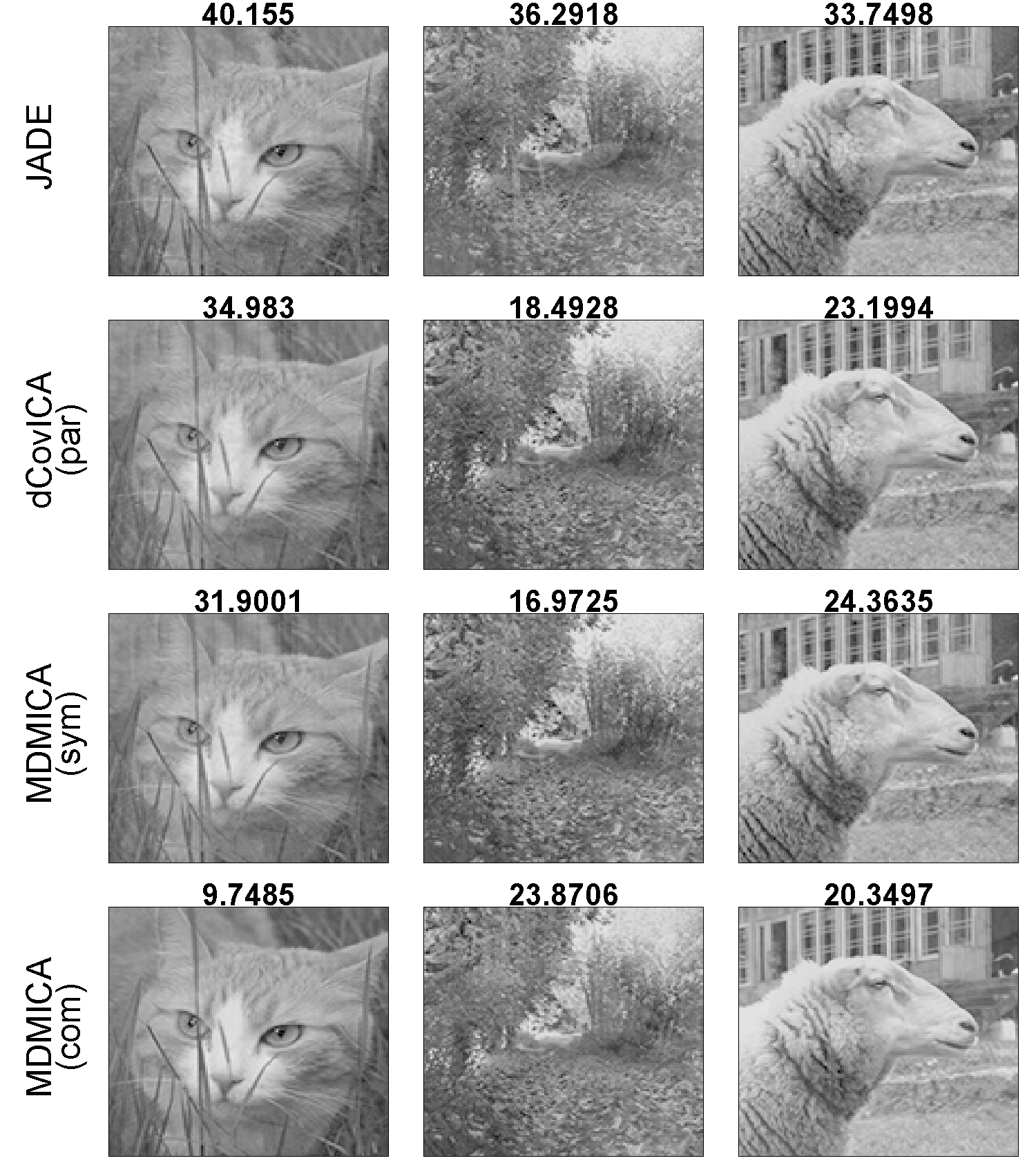}}
\caption{Recovered images with $p = 3$, $n = 130^2$ for the image data.
Each value on title is the Euclidean norm of the vectorized errors of the recovered image.
A signed permutation is applied to the images for illustration.}\label{fig5}
\end{center}
\vspace{-0.5cm}
\end{figure}

We estimate the intensity values $\widehat{\mathbf{S}}$
initialized from $10d$ points using LHS.
See Figures \ref{fig5} for the recovered images,
where the Euclidean norm of vectorized errors is computed to evaluate the estimation accuracy.
Indicated by the estimated images and errors, dCovICA and MDMICA outperforms JADE.
Moreover, MDMICA (com) achieves the best overall performance.


\section{CONCLUSION}\label{con}

Resorting to recently proposed mutual dependence measures
including MDMs in \citet{jin2017generalizing} and dHSIC in \citet{pfister2016kernel},
we generalize dCovICA in \citet{matteson2017independent} to a new ICA approach, MDMICA,
taking empirical dependence measures as an objective function for the estimation of ICs.
In addition, we study the asymptotic properties of MDMICA.

When solving the non-convex minimization problem to estimate ICs,
we apply LHS and BO to select a better initial point for the Newton-type local optimization method.


MDMICA achieves competitive results with JADE and dCovICA,
and outperforms FastICA and Infomax in numerical studies,
under different distributions and dimensions of ICs.
When the ICA model is misspecified, MDMICA decreases the mutual dependence between components via optimization,
while other approaches cannot and even increase it.
We illustrate the advantage of using multiple initial points from LHS and BO over a single initial point.


During the image recovery task from mixed image data,
MDMICA not only nicely recovers the true images,
but also achieves lower overall errors than other approaches,
which demonstrates the value of MDMICA in real data applications.




%


%
%


\renewcommand{\refname}{\normalsize{References}}
\bibliographystyle{abbrvnat}
\bibliography{Paper}

\end{document}